\documentclass[12pt,prd,aps,epsfig,floats,preprint,eqsecnum,superscriptaddress]{revtex4}
\usepackage{graphicx,amsmath}
\usepackage{subfigure}

\newcommand{\be}{\begin{equation}}
\newcommand{\ee}{\end{equation}}
\newcommand{\bea}{\begin{eqnarray}}
\newcommand{\eea}{\end{eqnarray}}
\newcommand{\beq}{\begin{equation}}
\newcommand{\eeq}{\end{equation}}
\newcommand{\beqa}{\begin{eqnarray}}
\newcommand{\eeqa}{\end{eqnarray}}
\newcommand{\no}{\nonumber}
\def\lsim{\mathrel{\rlap{\lower4pt\hbox{\hskip1pt$\sim$}}
     \raise1pt\hbox{$<$}}}         %less than or approx. symbol
\def\gsim{\mathrel{\rlap{\lower4pt\hbox{\hskip1pt$\sim$}}
     \raise1pt\hbox{$>$}}}         %greater than or approx. symbol

\begin{document}
%\draft

%\preprint{{\vbox{\hbox{}\hbox{}\hbox{}
%\hbox{WIS/??/08-Nov-DPP}
%\hbox{hep-ph/yymmnnn}}}}

\vspace*{.0cm}

\title{Testing minimal lepton flavor violation with\\
extra vector-like leptons at the LHC}

\author{Eilam Gross} \email{eilam.gross@weizmann.ac.il}
\affiliation{Department of Particle Physics and Astrophysics,
   Weizmann Institute of Science, Rehovot 76100, Israel}
\author{Daniel Grossman}\email{daniel.grossman@weizmann.ac.il}
\affiliation{Department of Particle Physics and Astrophysics,
  Weizmann Institute of Science, Rehovot 76100, Israel}
\author{Yosef Nir\footnote{The Amos de-Shalit chair of theoretical
    physics}}\email{yosef.nir,@weizmann.ac.il}
\affiliation{Department of Particle Physics and Astrophysics,
  Weizmann Institute of Science, Rehovot 76100, Israel}
\author{Ofer Vitells}\email{ofer.vitells@weizmann.ac.il}
\affiliation{Department of Particle Physics and Astrophysics,
   Weizmann Institute of Science, Rehovot 76100, Israel}
%\vspace*{1cm}}

\vspace*{1cm}
\begin{abstract}
  Models of minimal lepton flavor violation where the seesaw scale is
  higher than the relevant flavor scale predict that all lepton flavor
  violation is proportional to the charged lepton Yukawa matrix. If
  extra vector-like leptons are within the reach of the LHC, it will
  be possible to test the resulting predictions in ATLAS/CMS.
\end{abstract}
\maketitle

%%%%%%%%%%%%%%%%%%%%%%
%%%%%%%%%%%%%%%%%%%%%%
\section{Introduction}
\label{sec:int}
Measurements of flavor changing processes in meson decays are all in
agreement with the Standard Model (SM) predictions. Such a situation
is not expected if there is new physics at the TeV scale, unless its
flavor structure resembles that of the SM. The strongest suppression
of the new physics flavor effects would arise if all new flavor
couplings were proportional to the SM Yukawa couplings, $Y^U$ and
$Y^D$, an idea that became known as ``minimal flavor violation'' (MFV)
\cite{D'Ambrosio:2002ex,Hall:1990ac,Chivukula:1987py,Buras:2000dm,Kagan:2009bn}
and which applies, for example, in several known supersymmetric
models, such as gauge mediation.

As concerns the lepton sector, the fact that no flavor changing
neutral current (FCNC) decays of charged leptons have been observed
suggests that a similar principle -- minimal lepton flavor violation
(MLFV) -- might apply
\cite{Cirigliano:2005ck,Cirigliano:2006su,Cirigliano:2006nu,Branco:2006hz,Chen:2008qg}.
The existence of neutrino masses, however,
implies that there are at least two possible scenarios of MLFV.  It is
quite likely that the seesaw mechanism, involving heavy singlet
fermions with masses $m_N\gg m_Z$, is responsible for the generation of
the light neutrino masses. If the mass scale $m_N$ is lower than the
scale of flavor dynamics, then there could be three relevant flavor
violating matrices: The Yukawa matrix of the charged leptons $Y^E$,
the Yukawa matrix of the neutrinos $Y^N$, and the heavy neutrino mass
matrix $M_N$. If $m_N$ is higher than the scale of flavor dynamics,
then MLFV requires that all low energy flavor violating couplings are
proportional to $Y^E$. In this work, we use the term MLFV for the
latter scenario only.

While the high $p_T$ experiments at the LHC, ATLAS and CMS, have not
been constructed as flavor machines, the fact that they can identify
electrons and muons with high precision makes them potentially
powerful probes of lepton flavor physics. If new particles, with
masses within the reach of the LHC, decay into the SM charged leptons,
then ATLAS and CMS are uniquely capable of probing detailed features
of the new particles, which may be crucial in understanding the
underlying theory. This has been demonstrated for various
classes of supersymmetric models
\cite{Allanach:2008ib,Feng:2007ke,Feng:2009yq,Feng:2009bd,Buras:2009sg}.
(Implications of quark-related MFV for LHC phenomenology have also
been explored
\cite{Gresham:2007ri,Grossman:2007bd,Dittmaier:2007uw,Hiller:2008wp,Burgess:2009wm,Hiller:2009ii,Arnold:2009ay}.)

In this work, we focus on an extension of the SM where there are
heavy -- but still within the reach of the LHC -- vector-like
doublet-leptons. MLFV gives strong predictions concerning the spectrum
and the couplings of such new leptons. We analyze how, and to what
extent, ATLAS and CMS can test the MLFV hypothesis with such new
particles.

The plan of this paper goes as follows. In Section \ref{sec:the}
we present our theoretical framework. In Section \ref{sec:exp}
we describe the LHC phenomenology. In Section \ref{sec:mlfv} we
analyze the lessons concerning minimal flavor violation that can be
drawn from the ATLAS/CMS measurements.

%%%%%%%%%%%%%%%%%%%%%%
%%%%%%%%%%%%%%%%%%%%%%
\section{The theoretical framework}
\label{sec:the}
The SM leptons include the lepton $SU(2)$-doublets $L_L$ and the
charged lepton $SU(2)$-singlets $E_R$. We assume that, in addition to
the SM leptons, there exist vector-like leptons, $\chi_L$ and
$\chi_R$, which are $SU(2)$-doublets and carry hypercharge $-1/2$ (so
that the electric charges of the two members in each doublet are $0$
and $-1$). The most general Yukawa and mass terms of the leptonic
sector in this extended framework are the following:
\beq\label{eq:llep}
{\cal L}_{\rm leptons}=-Y^E_{ij}\overline{L_{L}^i}\phi E_{R}^j
-(m_2/v)Y^\chi_{ij}\overline{\chi_{L}^i}\phi E_{R}^j -M_2
X^{\chi}_{ij}\overline{\chi_{L}^i}\chi_{R}^j -M_1
X^{L}_{ij}\overline{L_{L}^i}\chi_{R}^j.
\eeq
where $v=\langle\phi\rangle$, and $m_2,M_1,M_2$ have dimension of
mass.  The first two terms are Yukawa couplings and the last two bare
mass terms.  We introduce the ratio $m_2/v$ into the second term for
later convenience. We assume that the electroweak symmetry breaking
parameters $v$ and $m_2$ are smaller than the electroweak symmetry
conserving ones, $M_1$ and $M_2$.

%%%%%%%%%%
\subsection{The models}
\label{sec:models}
To implement the MLFV principle, we need to assign the various fields
to representations of the lepton flavor symmetry
\beq
G_{\rm LF}=SU(3)_L\times SU(3)_E.
\eeq
By definition, the SM lepton fields are triplets of $G_{\rm LF}$:
\beq
L_L(3,1),\ \ \ E_R(1,3),
\eeq
and the SM charged lepton Yukawa matrix acts as a spurion
which breaks $G_{\rm LF}$:
\beq
Y^E(3,\bar3).
\eeq
We are free to assign the new fields, $\chi_{L,R}$ to whichever
$G_{\rm LF}$ representation that we wish. The assignment
determines the spectrum and the couplings of these fields. We are
interested, however, in models where the $\chi$ fields couple to SM
leptons.  The simplest choice for that is to put them in triplets of
$G_{\rm LF}$.  There are four different ways to do that, which are
given in Table \ref{tab:models}.  We call the four resulting models as
LE, LL, EE, and EL in an obvious correlation to the way that $\chi_L$
and $\chi_R$ transform under $SU(3)_L\times SU(3)_E$.

\begin{table}[t]
  \caption{The four models with $\chi_{L,R}$ in triplets of $G_{\rm
      LF}$. The columns under $\chi_{L,R}$ give the $SU(3)_L\times
  SU(3)_E$ presentations -- triplets (3) or singlets (1). The entries
  in the $Y^\chi,X^\chi$ and $X^L$ columns give the flavor structure
  of the leading contribution to each of these matrices. (There is an
  arbitrary overall coefficient in each entry, which we assume to be
  of order one.)}
\label{tab:models}
\begin{center}
\begin{tabular}{ccc|ccc} \hline\hline
  \rule{0pt}{1.2em}%
Model & $\chi_L$ & $\chi_R$ &\ $Y^\chi$\ &\ $X^\chi$\ &\ $X^L$\
\cr \hline
LE & (3,1) & (1,3) & $Y^E$ & $Y^E$ & 0 \cr
LL & (3,1) & (3,1) & $Y^E$ & ${\bf1}$ & 0 \cr
EE & (1,3) & (1,3) & ${\bf1}$ & ${\bf1}$ & $Y^E$ \cr
EL & (1,3) & (3,1) & ${\bf1}$ & $Y^{E\dagger}$ & ${\bf1}$ \cr
\hline\hline
\end{tabular}
\end{center}
\end{table}

MLFV requires that the Lagrangian terms be made of the
$L_L,E_R,\chi_L$ and $\chi_R$ fields and of $Y^E$ spurions in a
formally $G_{\rm LF}$ invariant way. By definition, this holds for the
$Y^E$ term in Eq. (\ref{eq:llep}).  On the other hand, the
$Y^\chi,X^\chi$ and $X^L$ involve the new fields, and consequently
their structure is different in one model from the other:
\begin{itemize}
\item The LE model: $Y^\chi,X^\chi$ and $X^L$ must
all transform as $(3,\bar3)$ and are therefore proportional to $Y^E$.
Note, however, that the $L_L$ and $\chi_L$ fields transform in the same
way under both $G_{\rm SM}$ and $G_{\rm LF}$. There is therefore freedom
in choosing a basis in the $(L_L,\chi_L)$ space. We choose this freedom
to make $X^L=0$, namely we define the $\chi_L$ fields as the three fields
that have bare mass terms.
\item The LL model: $Y^\chi,X^\chi$ and $X^L$ must
transform as $(3,\bar3)$, $(1+8,1)$ and $(1+8,1)$, respectively.
We thus have $Y^\chi\propto Y^E$ and $X^\chi\propto{\bf1}$ while,
again, we are free to choose a basis in the $(L_L,\chi_L)$ space
 where $X^L=0$.
\item The EE model: $Y^\chi,X^\chi$ and $X^L$ must
transform as $(1,1+8)$, $(1,1+8)$ and $(3,\bar3)$, respectively.
We thus have $Y^\chi\propto{\bf1}$, $X^\chi\propto{\bf1}$ and
$X^L\propto Y^E$.
\item The EL model: $Y^\chi,X^\chi$ and $X^L$ must
transform as $(1,1+8)$, $(\bar3,3)$ and $(1+8,1)$, respectively.
We thus have $Y^\chi\propto{\bf1}$, $X^\chi\propto Y^{E\dagger}$
and $X^L\propto{\bf1}$. This model does not give the correct mass
hierarchy for the SM charged leptons unless we fine-tune either
$m_2$ or $M_1$ to be negligibly small.  We thus do not consider model
EL any further.
\end{itemize}

%%%%%%%%%%
\subsection{Masses}
The charged lepton mass matrix is a $6\times6$ Dirac mass matrix.
The neutral lepton mass matrix is a $9\times9$ Majorana mass matrix.
To obtain the mass eigenvalues and the mixing parameters we
need to diagonalize these matrices. However, the hierarchies $m_2\ll
M_2$ and $y_\tau\ll1$ allow us to obtain the main features
straightforwardly. In particular, the spectrum of the heavy leptons is
either quasi-degenerate (models EE and LL) or hierarchical, with
hierarchy proportional to that of the light charged leptons (model
LE). In order that we have at least one heavy lepton within the reach
of the LHC, we take $M_2\lsim TeV$ for the quasi-degenerate models,
and $y_e M_2\lsim TeV$ ($M_2\sim10^5\ TeV$) for the hierarchical
model.

%%%%%%%%%%
\subsection{Decays}
The leading decay modes of the heavy leptons would be two body decays
into a light lepton and either the Higgs boson, or the $Z$-boson or
the $W$-boson. Since the only lepton flavor violating spurion is
$Y^E$, then, neglecting neutrino masses, there remains an exact lepton
flavor symmetry,
\beq
G_{\rm LF}\to U(1)_e\times U(1)_\mu\times U(1)_\tau.
\eeq
Each of the heavy lepton mass eigenstates thus decays into one, and
only one light lepton flavor. This is the strongest prediction of our
MLFV framework, and it provides the most crucial tests.

To find the relevant couplings of the heavy leptons, one has to obtain
the interaction terms in the heavy lepton mass basis. However, the
leading contributions and the most important features can again be
understood on the basis of a straightforward spurion analysis.
We first note that the decays will be dominantly into either the Higgs
boson $h$ or the longitudinal components of the vector bosons,
$\phi_3$ and $\phi_\pm$. Therefore, the decays are chirality changing.
Furthermore, the $\chi_R\to E_L\phi$ transitions involve
$SU(2)$-breaking and are therefore suppressed by $m_2/M_2$. On the
other hand, the $\chi_L\to E_R\phi$ transitions are
$SU(2)$-conserving, and therefore proportional to $m_2/v$ which, by
assumption, is of order one.

To proceed we note that the rotation from the interaction basis to the
mass basis involves small rotation angles. We can therefore extract
the leading flavor structure by analyzing the flavor eigenstates. The
$\chi_L\to E_R$ transitions depend on $(m_2/v)Y^\chi$. Examining Table
\ref{tab:models}, we learn that in models LE and LL it will be
flavor-suppressed as $Y^E$, while in the EE model, it is
unsuppressed by flavor parameters.

In any case, the strongest suppression factor that appears in our
framework is $(m_2/v)y_e\sim10^{-5}$. Thus, the longest-lived lepton
can have a decay width of order $10^{-11}$ its mass, which still gives
a lifetime shorter than $10^{-16}$ seconds. We conclude that all the
heavy leptons decay promptly. Among the TeV scale leptons, the
shortest-lived has a width of order $1/(8\pi)$ of its mass, still too
narrow to be measured. We conclude that there is no way to measure the
decay width of the heavy vector leptons of our MLFV models in ATLAS/CMS.

Finally, we note that the following relation between the various
leading decay rates holds to a good approximation:
\beq
\Gamma(\chi^-\to h\ell^-)=2\Gamma(\chi^-\to Z\ell^-)=2\Gamma(\chi^0\to
W^+\ell^-).
\eeq

%%%%%%%%%%
\subsection{Electroweak precision measurements}
The presence of new $SU(2)$-doublets and effects of $SU(2)$-breaking
in their spectrum modify the predictions for the electroweak precision
measurements and, in particular, the $S$, $T$ and $U$ parameters
\cite{Peskin:1991sw}.

Consider, for example, the $T$ parameter. The shift $\Delta T$ due to
new contributions is related to the small mass splittings between the
neutral and the charged members in the heavy $SU(2)$-doublets. The
mass of the $i$'th heavy lepton doublet is of order $M_2 X^\chi_i$,
while the mass splitting is of order $(m_2 Y^\chi_i)^2/(M_2
X^\chi_i)$. (As explained above, MLFV requires that the $Y^\chi$ and
$X^\chi$ matrices are diagonal.) We thus have
\beq
\Delta T(\chi_i)={\cal O}\left[\left(\frac{\Delta
      m_{\chi_i}}{m_Z}\right)^2\right]
\approx\frac{(m_2 Y^\chi_i)^4}{(m_Z M_2 X^\chi_i)^2}.
\eeq
Putting $m_2\sim m_Z$ and examining Table \ref{tab:models}, we obtain
\beq
\Delta T(\chi_i)\sim(m_2/M_2)^2\times \begin{cases}
    y_i^2 & {\rm LE} \cr
    y_i^4 & {\rm LL} \cr
    1 & {\rm EE} \end{cases}.
\eeq
Since $m_2/M_2\sim10^{-6}\ (10^{-1})$ for model LE (LL,EE), we have
\beq
\Delta T(\chi)\sim\begin{cases}
    10^{-12}y_\tau^2\sim10^{-16} & {\rm LE} \cr
    10^{-2}y_\tau^4\sim10^{-10} & {\rm LL} \cr
    10^{-2} & {\rm EE} \end{cases}.
\eeq

Exact calculations confirm these rough estimates. We made similar
calculations for $\Delta S(\chi)$ and $\Delta U(\chi)$. We find that
for $m_2/m({\rm lightest}\ \chi)\lsim0.1$, our models satisfy the
constraints from electroweak precision measurements.

%%%%%%%%%%%%%%%%%%%%%%
%%%%%%%%%%%%%%%%%%%%%%
\section{LHC phenomenology}
\label{sec:exp}
%
%%%%%%%%%%
\subsection{Production}
\begin{figure}[htb]
\begin{center}
%\begin{tabular}{cc}
%\includegraphics[keepaspectratio=true,width=7cm,height=6cm]{modelLeDrellYan.eps}&
\includegraphics[width=12cm]{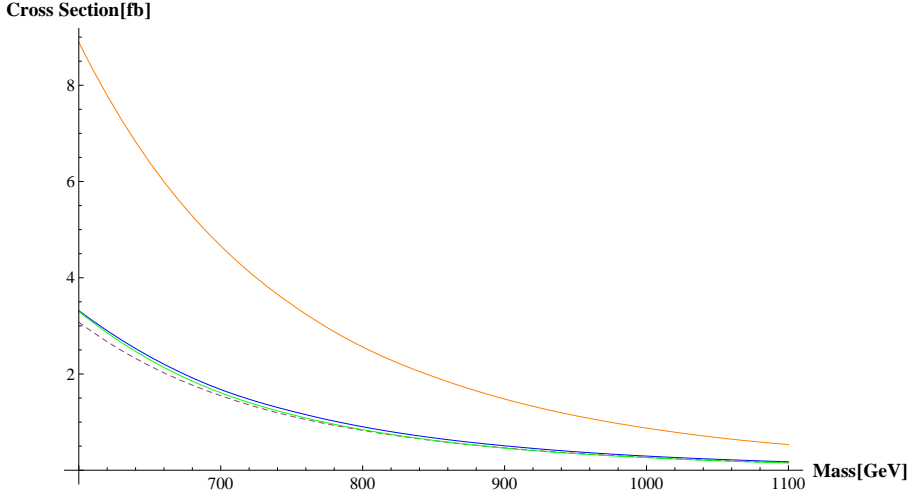}\\
%(a)&(b)
%\end{tabular}
\caption{Pair production cross sections at the LHC as a function of
  heavy lepton mass: $pp\rightarrow\chi^-\chi^+$ (solid blue),
  $pp\rightarrow \chi^0\bar{\chi}^0$ (dashed brown),
  $pp\to\chi^+\chi^0$ (solid orange) and $pp\to\chi^-\bar{\chi}^0$
  (solid green). The cross sections are given for a single heavy generation. }
\label{fig:productioncs}
\end{center}
\end{figure}

Since the heavy leptons are $SU(2)$-doublets, the main production
mechanism at the LHC will be $q\overline q^\prime\to
\chi\overline\chi$ via electroweak interactions. The production rate
is model independent. It is suppressed by the electroweak gauge
couplings, but not by any flavor factors. The most significant
process involves an intermediate $W^+$-boson, producing a heavy
charged lepton along with a heavy neutral lepton, $u\bar d\to
\chi^+\chi^0$. The second most important process is Drell-Yan
production involving an intermediate photon or $Z^0$-boson, $q\bar
q\to\chi^+\chi^-$. The production cross sections for a single
generation of vector-like heavy leptons are shown in
Fig.~\ref{fig:productioncs}. The simulation was
done using MadGraph v4 \cite{Alwall:2007st} with default cut values
at $E_{\rm cm}=14$ TeV and using CTEQ6L1 parton distribution
functions \cite{Kretzer:2003it}.

There are two points that we need to emphasize:
\begin{enumerate}
\item Within the MLFV framework, the production is
  always of a same flavor pair, {\it i.e.}
  $\chi_i\overline\chi_i$ (and not $\chi_i\bar\chi_j$ with $i\neq j$).
\item Since the coupling of heavy and light leptons is suppressed by
${\cal O}(v/M_2)$,  single heavy lepton production is negligible.
\end{enumerate}

%
%%%%%%%%%%
\subsection{Signature}
Most studies of heavy vector-like leptons assume no new
Yukawa interaction, so that the neutral heavy leptons are stable. This
improves the possibility of detection and allows for a variety of
detection strategies \cite{Allanach:2001sd} with an LHC mass reach of
$\sim 1\;TeV$. In our case, however, the heavy leptons decay
to SM leptons and electroweak gauge bosons or Higgs particles, leading
to final states with multiple leptons and light jets. In the case of a
light Higgs decaying predominantly into $b\bar{b}$ it is also possible
to have heavy $b$ jets, otherwise the Higgs decays into pairs of
electroweak gauge bosons allowing for many particles in the final state.
Although the decay products described above seem complicated, the lack
of final state neutrinos (except from $W$ and $Z$ decays) allows for a
detection strategy based on reconstruction of the heavy lepton mass.

\begin{figure}[htb]
\begin{center}
\includegraphics[width=10cm]{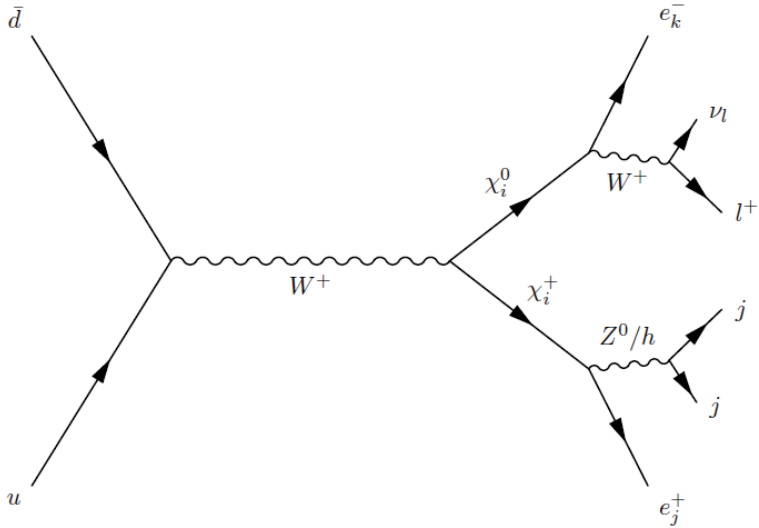}\\
\caption{The leading heavy leptons pair-production process, and the decay modes that we use for detection.}
\label{fig:dia}
\end{center}
\end{figure}

The process that we are looking at is
\beqa
pp&\to& \chi^+\chi^0,\\
&&\chi^+\to \ell_1^+ Z^0/h^0,\ \ \ Z^0/h^0\to{\rm jets},\no\\
&&\chi^0\to \ell_2^\mp W^\pm,\ \ \ W^\pm\to\nu\ell_3^\pm,\no
\eeqa
where $\ell$ stands for $e$ or $\mu$. The relevant diagram
is shown in Fig. \ref{fig:dia}. The main signature that we are looking
at is thus that of three isolated high $p_T$ leptons.

The process
\beq\label{NN}
pp\to\chi^0\bar\chi^0\to W^+W^-\ell^+\ell^-,
\eeq
where one of the $W$-bosons decays leptonically and the other decays
hadronically leads to the same final state, but it contributes at much
lower rate.

%%%%%%%%%%%%%
\subsection{Event selection}
The final state we are considering has a clean signature of three
isolated high $p_T$ leptons. Standard model processes with such a
final state are rare; the dominant sources are $t\bar t$ pairs with an
associated production of a $W/Z$ boson, as well as di-boson
production, $WZ$ and $ZZ$.  Since most of these processes involve a
leptonic $Z$ decay, they can be efficiently suppressed by imposing a
$Z$-veto, i.e. the requirement that no opposite-sign lepton pair is
present in the event with invariant mass close to that of the $Z$
boson. We have also considered as possible backgrounds $Zb\bar b$ and
di-lepton $t\bar t$, where additional leptons may be produced by the
decay of $B$-mesons in the $b$-jets. All signal and background samples
for this study were generated with MadGraph \cite{Alwall:2007st} at
$E_{\rm cm}=14$ TeV, with showering and hadronization done by PYTHIA
\cite{Sjostrand:2006za}, and detector effects simulated with the PGS
fast simulation package \cite{pgs}.

\begin{figure}[h!]
\begin{center}
\subfigure[]{\includegraphics[width=8cm]{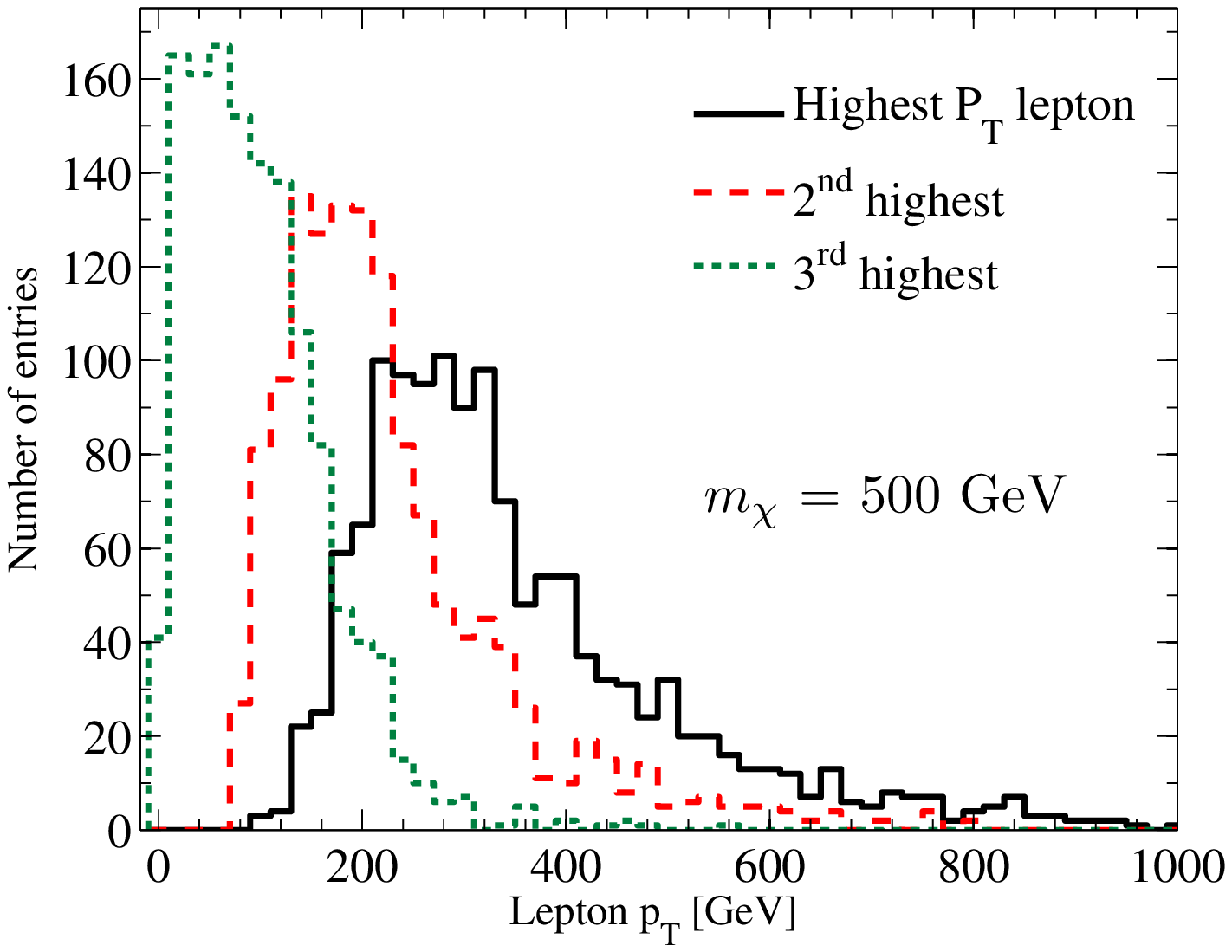}}
\subfigure[]{\includegraphics[width=8cm]{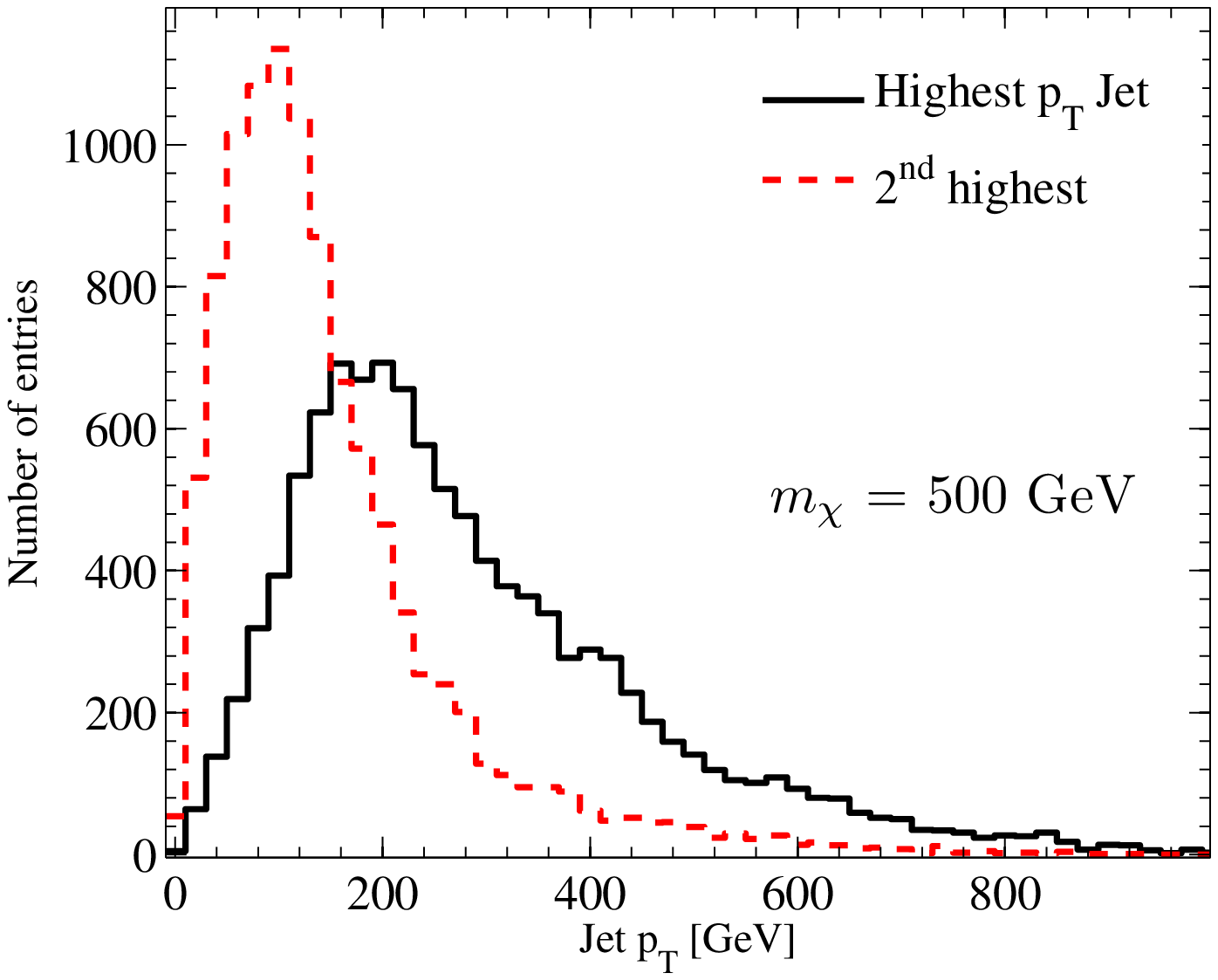}} \caption{
Distributions of transverse momenta in a signal sample corresponding
to $m_\chi=500$ GeV for (a) the three final charged leptons, and (b)
the two leading final jets.}
\label{fig:pt_dist}
\end{center}
\end{figure}

Fig. \ref{fig:pt_dist} shows the transverse momentum
distributions of the reconstructed leptons and the two leading jets,
for a signal sample with $m_{\chi}=500$ GeV.

Taking these distributions into consideration, we applied the
following selection cuts:
\begin{enumerate}
\item Exactly three isolated leptons, not all same-sign, with $p_T>25$
  GeV, of which at least two have $p_T>80$ GeV;
\item At least two jets with $p_T>25$ GeV or one jet with $p_T>50$ GeV;
\item The $Z$-veto is applied by requiring that $|m_{\ell^+\ell^-}-m_Z|>25$ GeV.
\end{enumerate}

Isolation cuts for electrons were applied by the default PGS
reconstruction algorithm.
The isolation cuts for muons are defined as follows:
\begin{enumerate}
\item The summed transverse momentum in $\Delta R=0.4$ cone around the
muon (excluding the muon itself) is $<5$ GeV;
\item The ratio of transverse energy in a grid of $3\times3$ calorimeter
cells around the muon (including the muons cell) to the transverse
momentum of the muon is $<0.1$.
\end{enumerate}

Table \ref{tab:select} presents the numbers
of events passing the selection criteria (in fb). The signal corresponds
to model LL, where there are three quasi-degenerate heavy leptons.

\begin{table}[h!]
  \caption{Number of events corresponding  to integrated luminosity of 1 fb$^{-1}$:
  total (first column); passing the selection criteria before imposing
  the $Z$ veto (second column); passing all cuts (third column). The last column
   reports the number of events that we generated for our simulation.
  The signal sample corresponds to three heavy lepton generations with $m_\chi=500$ GeV.
   It includes all production and decay modes ({\it i.e.}
   $\chi^0\bar\chi^0,\chi^+\chi^-,\chi^\pm\chi^0$), and the branching
   ratio refers to a three lepton final state. We use $m_h=120$ GeV.
   }
   \label{tab:select}
\begin{center}
\begin{tabular}{l c ||c|c|c||c} \hline\hline
  \rule{0pt}{1.2em}%
Process & $W/Z$ Decay & $\sigma \times\mathcal{BR}$\ [fb] &
Selection [fb]& $Z$ veto [fb] & Generated events \cr \hline
$t\bar t Z$ & $Z \to \ell^+\ell^-$ & 155.7 & 2.19 & 0.052 & 32.7K \cr
$t\bar t W$ & $3W \to 3(\ell\nu)$ & 13.95 & 0.174 & 0.139 & 23.7K  \cr
$ZZ$ & $2Z \to 2(\ell^+\ell^-)$ & 71.6 & 0.632 & 0.004 & 10K \cr
$WZ$ & $Z \to \ell^+\ell^-$, $W \to\ell\nu$ & 157 & 0.471 & $< 0.016$ & 10K \cr
$t\bar t$ & $2W\to2(\ell\nu)$ & 33329 & 0.054 & 0.018 & 1.8M \cr
$b\bar b Z$ & $Z \to\ell^+\ell^-$ & 60000 & 0.027 & $<0.027$ & 2.3M \cr \hline
Signal & & 19.0 & 12.8 & 12.0 & 25K \cr
\hline\hline
\end{tabular}
\end{center}
\end{table}

%%%%%%%%%%
\subsection{Reconstruction}
\begin{figure}[h!]
\begin{center}
\includegraphics[width=10cm]{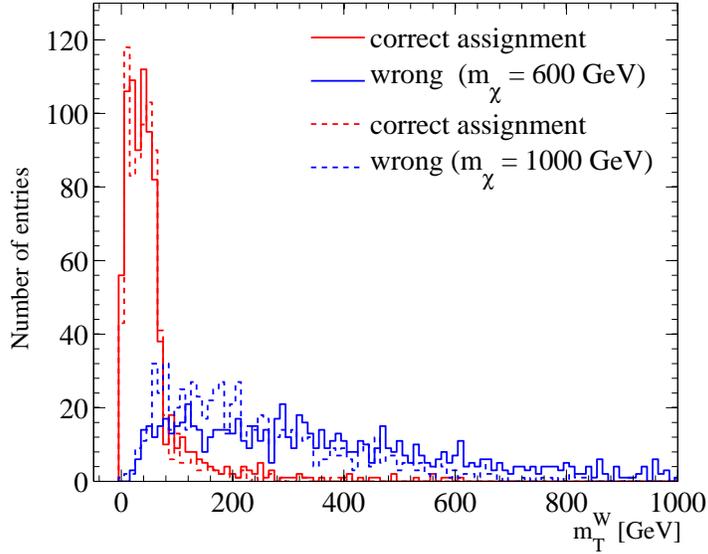}
\caption{Distributions of the $W$ transverse mass with correct (red)
and wrong (blue) lepton assignments, for heavy lepton masses of 600
(solid) and 1000 (dotted) GeV.} \label{fig:mtw}
\end{center}
\end{figure}

Reconstruction of the heavy lepton mass requires the identification of
the SM lepton originating from the $W$ decay. One possibility is immediately
ruled out, since this lepton can only be one
of the two leptons which have the same sign. We have calculated the
transverse mass of the $W$ for both of those leptons: \beq
(m_T^W)^2=2p_T^\ell p_T^{\rm miss}(1-\cos\phi_{\ell,{\rm miss}}),
\eeq
The distributions of $m_T^W$ are shown in Fig. \ref{fig:mtw}, for the correct
and for the wrong lepton assignments. The combination that yields the lower
value was designated as the $W$ decay product. The correct lepton configuration
was selected with this procedure at about 93\% of the events.

The two remaining opposite sign leptons, assumed to be produced
directly by the heavy lepton-pair decays, were then assigned to the
charged and neutral lepton decays according to their charges. Note
that the above reconstruction procedure equally applies to events
with a heavy neutral lepton pair and the same final state (\ref{NN}).

The transverse mass of the heavy neutral lepton was calculated
according to \beq (m_T^{\chi^0})^2=m_{\ell^+\ell^-}^2+2(E_T^{\chi^0}
p_T^{\rm miss}- {\bf p}_T^{\chi^0}\cdot {\bf p}_T^{\rm miss}), \eeq
where ${\bf p}_T^{\chi^0}={\bf p}_T^{\ell^+}+{\bf p}_T^{\ell^-}$,
with $\ell^+$ and $\ell^-$ the two leptons associated with the
$\chi^0\to W^+\ell^-\to\nu\ell^+\ell^-$ decay, and
$E_T^{\chi^0}=\sqrt{m_{\ell^+\ell^-}^2+|{\bf p}_T^{\chi^0}|^2}$.

\begin{figure}[h!]
\begin{center}
\subfigure[]{\includegraphics[width=8cm]{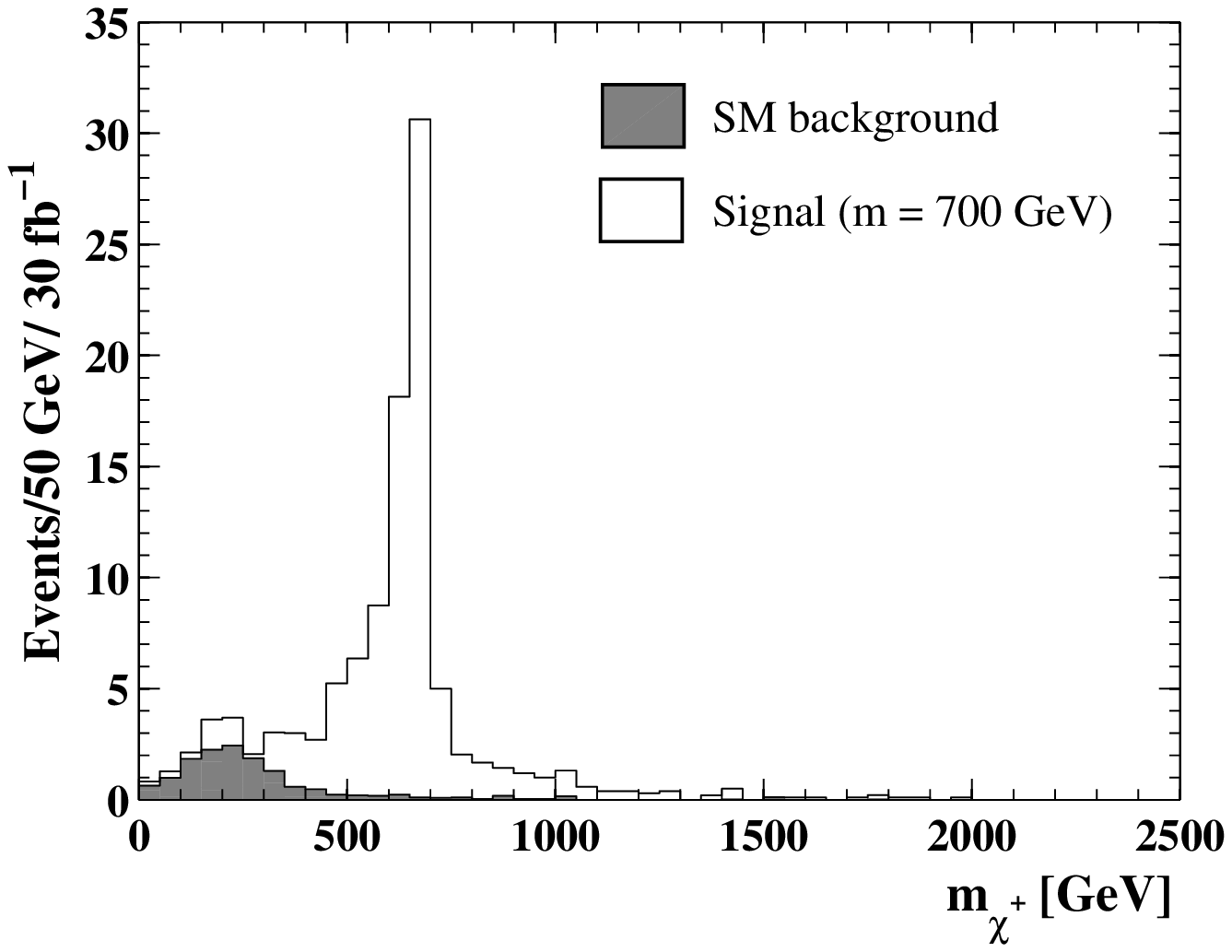}}
\subfigure[]{\includegraphics[width=8cm]{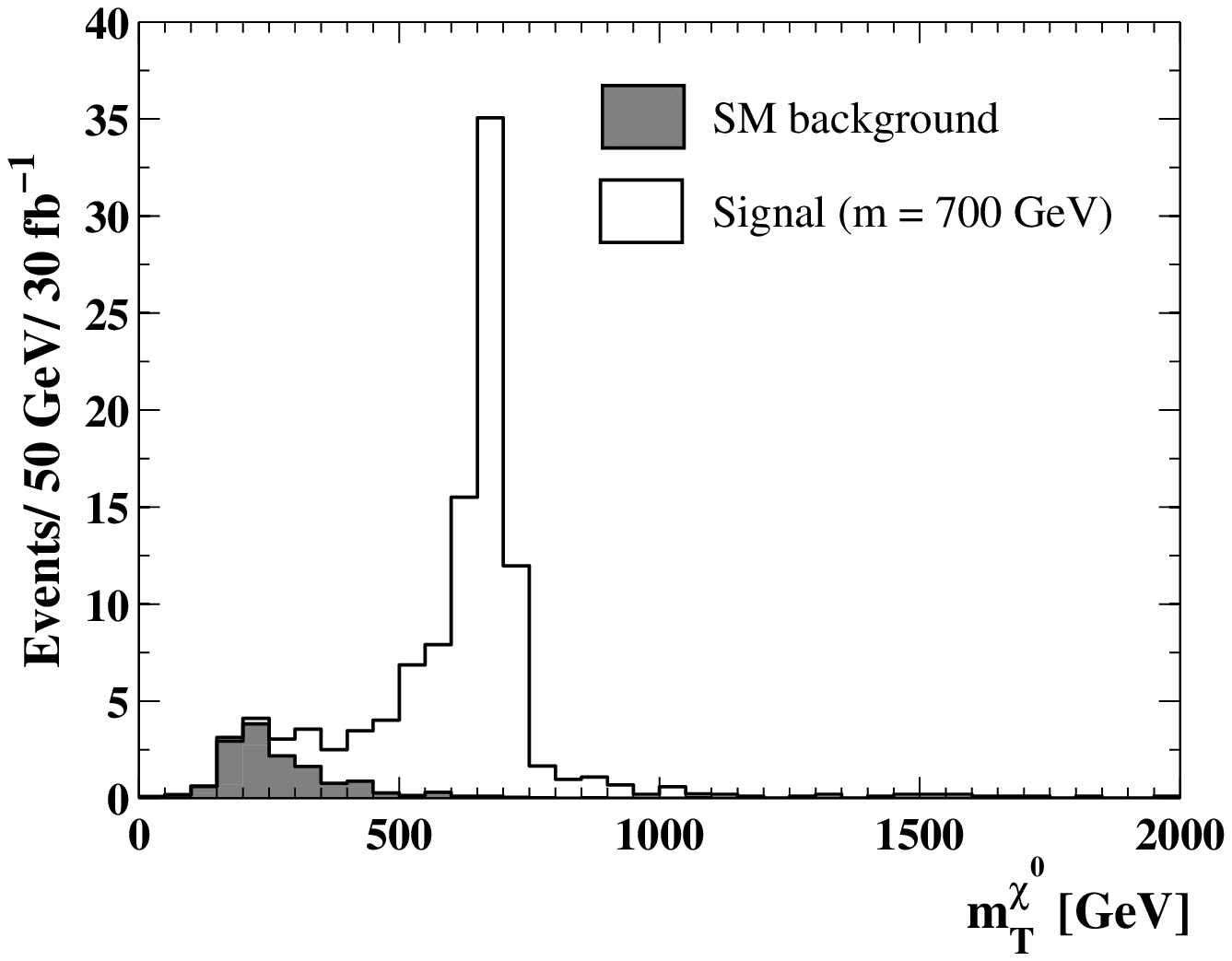}}
\caption{(a) Reconstructed invariant mass of the heavy charged
lepton. (b) Reconstructed transverse mass of the neutral heavy
lepton.} \label{fig:histo}
\end{center}
\end{figure}

The invariant mass of the heavy charged lepton was reconstructed from
the momenta of the two highest $p_T$ jets in the event and the
lepton that has opposite charge to that of the $W$:
\beq\label{eq:mchi} m_{\chi^\pm}^2=(p_{j1}+p_{j2}+p_\ell)^2. \eeq If
the $Z/h$ is highly boosted, it can be reconstructed as a single
jet. Therefore in the case that there is only a single reconstructed jet
in the event with $p_T>50$ GeV, $p_{j2}$ is omitted from
(\ref{eq:mchi}). The distributions of the reconstructed
$m_{\chi^\pm}$ and $m_T^{\chi^0}$ are shown in Fig.
\ref{fig:histo}, for $m_\chi=700$ GeV.

%
%%%%%%%%%%
\subsection{Obtaining flavor constraints}
\label{sec:flavor}
We focus here on model LL, which has three quasi-degenerate heavy
leptons, each decaying to one of the light lepton flavors,
$e,\mu,\tau$. Events are classified by the flavor of the two leptons
associated with the heavy pair decay. We are interested in $N_{ij}$,
the observed numbers of events in each flavor composition
$\ell_i^\pm\ell_j^\mp$. The MLFV prediction is that
\beqa\label{eq:mfvprea}
N_{ee}&=&N_{\mu\mu}=N_{\tau\tau},\no\\
N_{e\mu}&=&N_{e\tau}=N_{\mu\tau}=0.
\eeqa
Our analysis allows us to test two of these predictions, namely
\beq\label{eq:mfvpreb}
N_{ee}=N_{\mu\mu},\ \ \ N_{e\mu}=0.
\eeq

\begin{figure}[h!]
\begin{center}
\subfigure[]{\includegraphics[width=5cm]{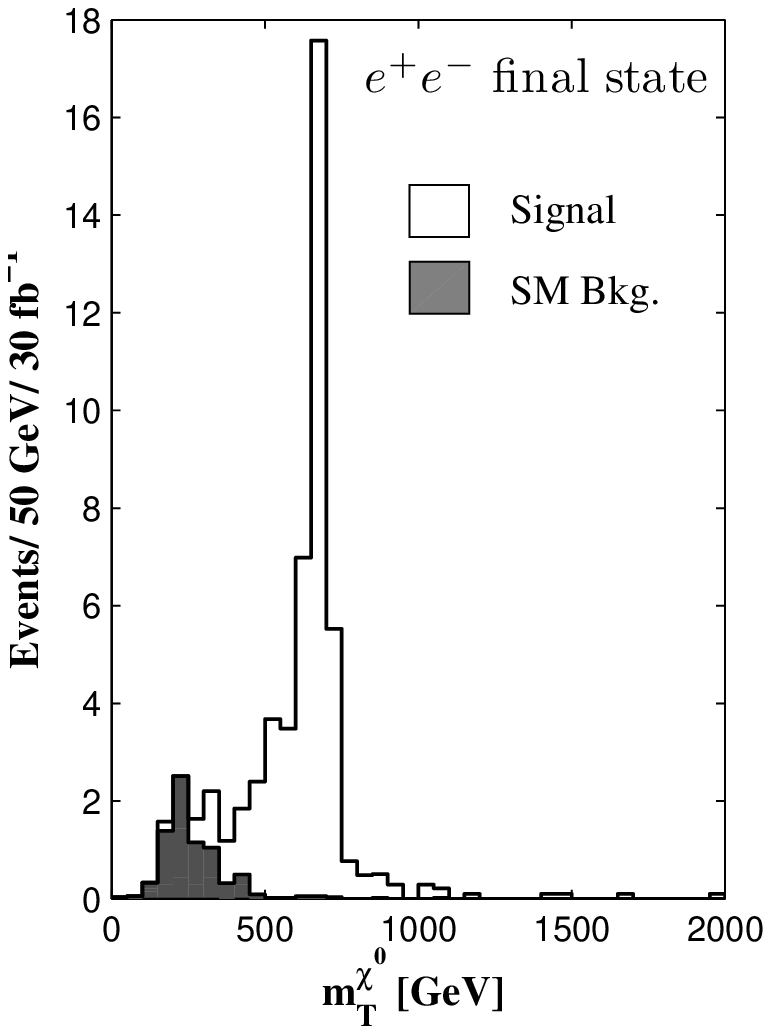}}
\subfigure[]{\includegraphics[width=5cm]{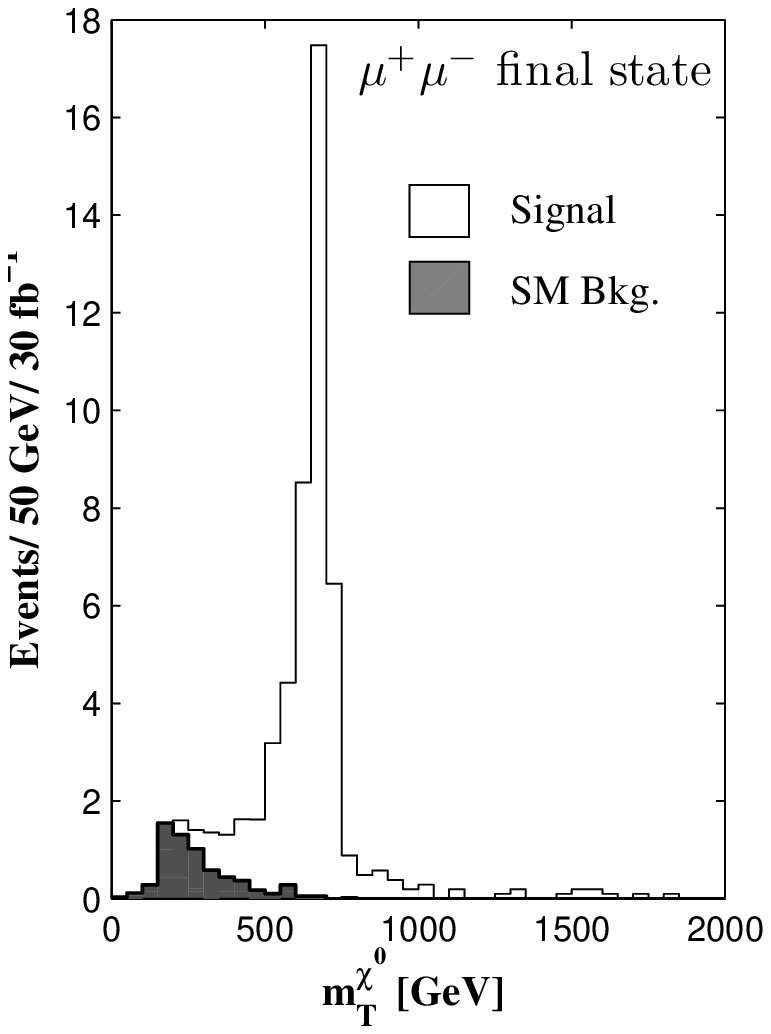}}
\subfigure[]{\includegraphics[width=5cm]{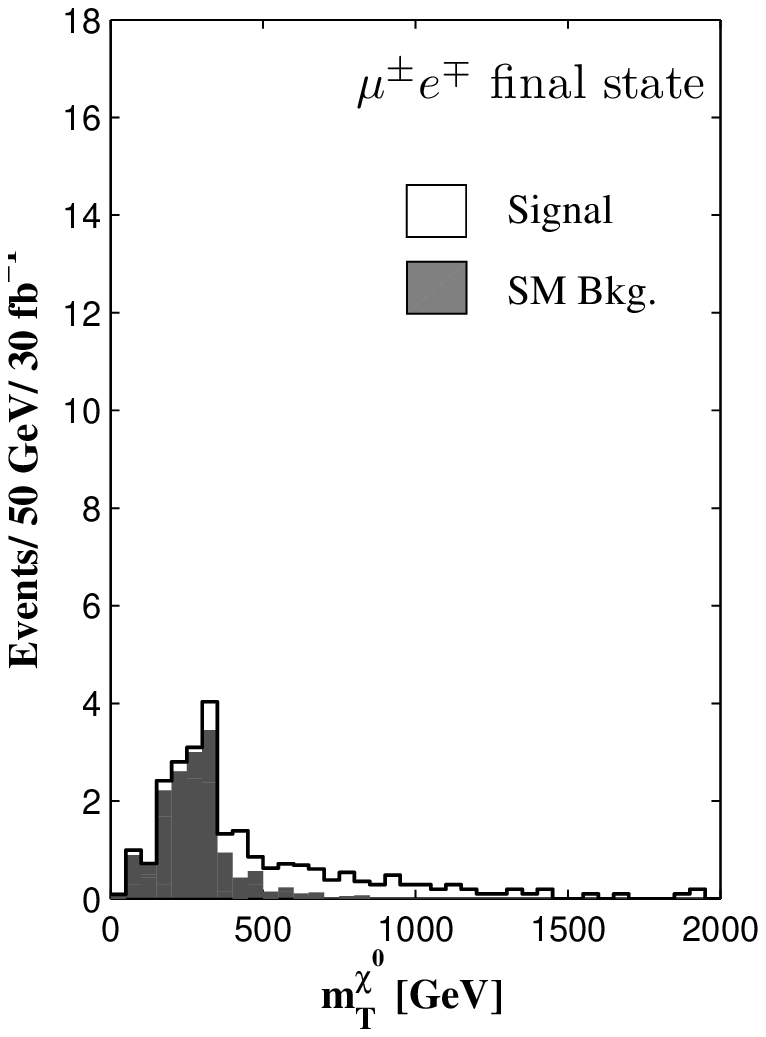}}
\caption{Number of events as a function of the reconstructed
transverse mass $m_T^0$, for $m_\chi = 700$ GeV and with an
integrated luminosity of $30$ fb$^{-1}$, for the different flavor
final states: (a) $ee$, (b) $\mu\mu$, and (c) $e\mu$.}
\label{fig:histo3}
\end{center}
\end{figure}

For the flavored cross section ratio estimates, we considered events
within a window of 150 GeV around the mass peak of both
$m_{\chi^\pm}$ and $m_T^{\chi^0}$. As is evident in Fig.
\ref{fig:histo}, standard model background in this region is
negligible. In Fig. \ref{fig:histo3}, the reconstructed
transverse mass $m_T^{\chi^0}$ is shown separately for the three
different flavor compositions, $e^+e^-$, $\mu^+\mu^-$ and
$e^\pm\mu^\mp$.  Ideally, there should be no events in
the $e\mu$ final state. In practice, however, a small number of the
signal events are reconstructed as such, mostly due to
misclassification of leptons in the event. Another possible
source of contamination are $\tau$ pairs, decaying to $e$ and $\mu$,
however this contribution was found to be negligible.

\begin{figure}[h!]
\begin{center}
%\begin{tabular}{cc}
%\includegraphics[keepaspectratio=true,width=7cm,height=6cm]{ee_mm_ratio.eps}&
\includegraphics[keepaspectratio=true,width=12cm,height=12cm]{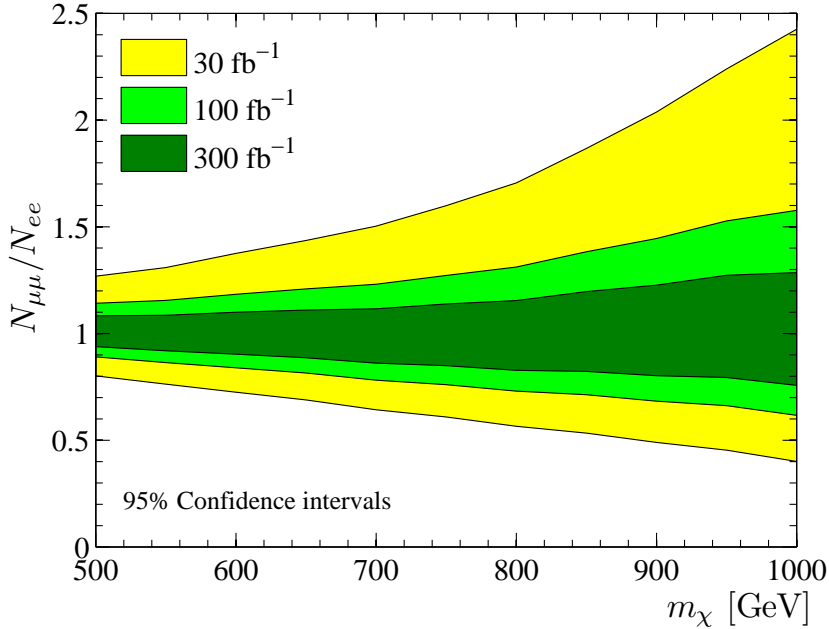}\\
%(a)&(b)
%\end{tabular}
\caption{The power of the LHC experiments to constrain the ratio
  $N_{\mu^+\mu^-}/N_{e^+e^-}$ (for models where this ratio is unity) as a
  function of the heavy
  lepton mass. A value outside the colored area can be rejected
  at 95\% CL for the corresponding integrated luminosity (dark green,
  light green, yellow for, respectively, 300, 100, 30 fb$^{-1}$).  }
\label{fig:eemm}
\end{center}
\end{figure}

\begin{figure}[h!]
\begin{center}
%\begin{tabular}{cc}
%\includegraphics[keepaspectratio=true,width=7cm,height=6cm]{ee_mm_ratio.eps}&
\resizebox{\textwidth}{!}{\rotatebox{-90}{
\includegraphics[width=10cm]{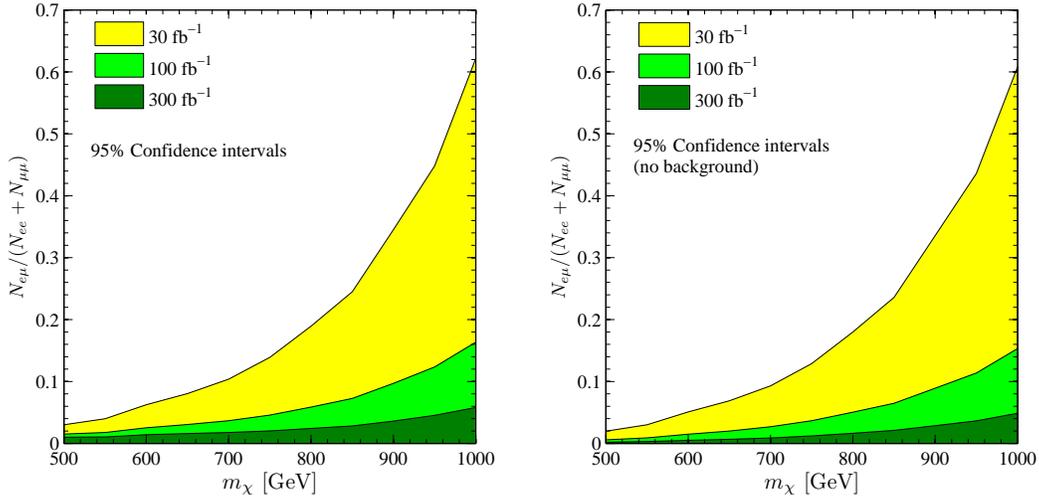}
} }
%(a)&(b)
%\end{tabular}
\caption{Right: The power of the LHC experiments to constrain the
ratio $N_{e\mu}/(N_{ee}+N_{\mu\mu})$ (for models where this ratio
is zero) as a function of the heavy lepton mass. A value above the
colored region can be rejected
  at 95\% CL for the corresponding integrated luminosity (dark green,
  light green, yellow for, respectively, 300, 100, 30 fb$^{-1}$).
  Left: similar limits for an ideal scenario in which there is
  no background, such that the uncertainty is purely statistical.  }
\label{fig:em}
\end{center}
\end{figure}

To set limits on the ratios of different flavor final states we
have treated the observed number of events of each category as
independent Poisson variables. In such a case, the exact confidence
intervals at a confidence level $1-\alpha$ are given by the
following formula \cite{James:1980my}:
\beqa
B_{L,U}&=&A_{L,U}/(1-A_{L,U}),\\
A_L&=&F^{-1}(\alpha/2;n_2,n_1+1),\no\\
A_U&=&F^{-1}(1-\alpha/2;n_2+1,n_1),\no
\eeqa
where $F(p;a,b)$ is the
cumulative distribution function of a Beta distribution with
parameters $a$ and $b$, at a value $p$, and $n_{1,2}$ are the
observed numbers of events. $B_{L}$ and $B_U$ are the lower and upper
bounds, respectively. The results are shown in Figs. \ref{fig:eemm}
and \ref{fig:em}. For the ratio $N_{e\mu}/(N_{ee}+N_{\mu\mu})$, the
presence of small number of signal events in the $\mu^\pm e^\mp$
final state, due to misclassification, slightly weakens the obtained
upper limit. This effect however is very small; to demonstrate this
we also consider an ``ideal'' scenario in which the number of
observed events $N_{e\mu}$ is set exactly to zero, as would be
expected in our model in case of perfect reconstruction and no
backgrounds. Those ideal limits are also shown in Fig.
\ref{fig:em}. For example, for a heavy lepton mass of $m_{\chi}=500$
GeV and with 30 fb$^{-1}$, the upper bound is only degraded due to
backgrounds from approximately 0.02 to 0.03.

%\begin{figure}
%  \resizebox{\textwidth}{!}{\rotatebox{-90}{\includegraphics{figure.eps}}}
%  \caption[short caption for table of figures]{Long Caption to appear
%  with the figure.  \label{figurelabel} }
%\end{figure}

The obtained limits are given for the ratios of
observed number of events. Within a realistic experimental
environment one would have to take into account the different
detection efficiencies of electrons {\it vs.} muons (which are
approximately equal in PGS). The difference in energy resolution
might also play a role. This effect is expected, however, to be
very small, since the resolutions of the reconstructed masses
($m_T^{\chi^0}$ and $m_{\chi^\pm}$) are mostly driven by
the energy resolution of jets. The ratio of reconstruction
efficiencies of electrons {\it vs.} muons could be measured to a very high
accuracy by comparing {\it e.g.} $Z \to e^+e^-$ to $Z \to
\mu^+\mu^-$ events. With $\mathcal{O}(10^5)$ such events expected per
$1\;{\rm fb}^{-1}$, the attainable uncertainty of the efficiency ratio is
expected to be negligible for our purposes. Thus, while a detailed
study of such experimental effects is beyond the scope of this work,
we expect the results presented here to be robust.

%%%%%%%%%%%%%%%%%%%%%%
%%%%%%%%%%%%%%%%%%%%%%
\section{Implications for MLFV}
\label{sec:mlfv}
The models presented in Section \ref{sec:models} demonstrate
that there could be a variety of mass spectra and couplings that
are consistent with the principle of MLFV. In particular,
\begin{itemize}
\item The mass spectrum can be either quasi-degenerate or hierarchical.
In the first case, we may have three heavy leptons within the reach of the
LHC, in the latter only one.
\item The couplings of the heavy vector-like leptons to the light, chiral
ones can be either universal or hierarchical. While this has an effect on
the lifetimes (which cannot be measured), it does not affect the overall
number of events in each flavor.
\end{itemize}

There is, however, one feature that that is common to all our MLFV models:
\begin{itemize}
\item The couplings of the heavy vector-like leptons to the light, chiral
ones are flavor-diagonal. In other words, we can describe the heavy lepton
mass eigenstates as, approximately, heavy electron, muon and tau.
\end{itemize}

We are able to test the diagonality of the couplings in two independent ways,
which are described in Section \ref{sec:flavor}. First, the comparison of the
number of $e^+e^-$ events to the number of $\mu^+\mu^-$ events, where the MLFV
prediction, for the case that both types of
events are observed, is one. (The other possibility, in case of hierarchical
spectrum, is that there are only $e^+e^-$ events.) As can be seen from Fig.
\ref{fig:eemm}, with 300 fb$^{-1}$ and $m_\chi\sim500$ GeV, this prediction
can be tested with an accuracy of order ten percent. With 30 fb$^{-1}$ and
$m_\chi\sim$ TeV, this prediction can be tested to within a factor of 2.5.

Second, we can search for $e\mu$ events which, according to MLFV, should not
be present. As can be seen from Fig.
\ref{fig:em}, with 300 fb$^{-1}$ and $m_\chi\sim500$ GeV, the ratio between
the flavor non-diagonal and flavor-diagonal events can be constrained to
lie below the percent level. With 30 fb$^{-1}$ and
$m_\chi\sim$ TeV, the bound is of order 0.6.

Low energy searches for flavor changing neutral current decays, such as
$\mu\to e\gamma$, put strong constraints on the product of the mass splitting
and the mixing angle between the heavy leptons.  Regardless of the strength of
such low energy constraints, ATLAS/CMS can provide flavor information that is not
available from low energy data. In particular, the $e\mu$-test will constrain
the mixing angle in the heavy sector for any finite mass splitting.

When ATLAS and CMS experiments collect enough data, they will also be
able to understand in more detail their capabilities in identifying
tau-leptons. It will become possible then to test also all tau-related
predictions of Eq. (\ref{eq:mfvprea}). While the experimental accuracy
of these measurements is expected to be poorer than the tests of
Eq. (\ref{eq:mfvpreb}), it may well be that violations of
MLFV predictions are larger when tau-leptons are involved.

The analysis proposed in this paper will become much easier if, in addition to
the charged heavy leptons, there exists a $Z^\prime$-boson that is light
enough to be produced at the LHC and heavy enough to decay into a
$\chi\bar\chi$ pair. Indeed, such a scenario, with stable heavy
leptons, was described in Ref.~\cite{Bauer:2009cc} as a scenario that
can be probed by a low energy and low luminosity initial LHC data set,
and which is not ruled out by the Tevatron and other measurements. In
such a case, we expect an ${\cal O}(16\pi^2)$ enhancement in the
number of signal events. It would mean that some informative (though
rough) flavor measurements will be possible with as little as few
hundreds of pb$^{-1}$ of integrated luminosity.

%%%%%%%%%%
\section*{Acknowledgements}
E.G. is obliged to the Benoziyo center for High Energy Physics, to the
Israeli Science Foundation (ISF), the Minerva Gesellschaft and the
German Israeli Foundation (GIF) for supporting this work.  The work of
Y.N. is supported by the Israel Science Foundation (ISF) under grant
No.~377/07, by the German-Israeli foundation for scientific research
and development (GIF), and by the United States-Israel Binational
Science Foundation (BSF), Jerusalem, Israel.

%%%%%%%%%%

\end{document}